\documentclass{aa}
\usepackage{epsfig}
\input{epsf}
\begin{document}
\title{Three-dimensional stability of the solar tachocline}
\titlerunning{Tachocline stability}
\author{Rainer Arlt\inst{1}\and Aniket Sule\inst{1}\and G\"unther R\"udiger\inst{1,2}}
\authorrunning{Rainer Arlt, Aniket Sule, G\"unther R\"udiger}
\institute{Astrophysikalisches Institut Potsdam, An der Sternwarte 16, Germany
\and Isaac Newton Institute for Mathematical Sciences, 20 Clarkson Road, Cambridge, CB3\,0EH, UK}

\date{\today}

\abstract{The three-dimensional, hydrodynamic stability of the solar 
tachocline is investigated based on a rotation profile as a function 
of both latitude and radius. By varying the amplitude of the latitudinal 
differential rotation, we find linear stability limits at various Reynolds 
numbers by numerical computations. We repeated the computations with different 
latitudinal and radial dependences of the angular velocity. The stability 
limits are all higher than those previously found from two-dimensional 
approximations and higher than the shear expected in the Sun. It is 
concluded that any part of the tachocline which is radiative is 
hydrodynamically stable against small perturbations.
}

\maketitle

\section{Motivation}
Helioseismology has provided us with information on the internal rotation
of the Sun. The angular velocity depends on latitude as well as radius.
The dependence is mainly a latitudinal in the bulk of the convection
zone, whereas the solar radiative core rotates nearly uniform with an
angular velocity of the convection zone at about $30^\circ$ latitude.
The transition from the differential rotation in the convection zone
to the uniform rotation in the core is thin -- probably thinner than
5\% of the solar radius -- and is called the tachocline.

The convection zone is thermally overcritical and the stability of
shear flows is not an issue there. The shear in the tachocline however
is latitudinal and radial and may be subject to shear-flow instabilities.
If the tachocline is turned into a turbulent layer, a problem arises
with the mixing of elements, most notably with Lithium which will be
destroyed in nuclear fusion up to 0.68 solar radii ($R_\odot$), 
just below the convection zone. A turbulent tachocline mixing the 
Lithium into its fusion zone would contradict the observed relatively 
high Lithium abundance at the surface of the Sun.

The situation in the solar tachocline was described by Spiegel \& Zahn
(1992) as exhibiting horizontal turbulence, excited by hydrodynamic
shear-flow instability. It was argued that the two-dimensional
turbulence provides angular momentum transport but inhibits too
strong mixing of Lithium below $0.68R_\odot$.

The hydrodynamic stability of the latitudinal shear in the tachocline
was studied by Watson (1981). The dependence on colatitude $\theta$
was $\Omega = \Omega_{\rm eq}(1-\alpha_2 \cos^2\theta)$, where
$\Omega_{\rm eq}$ is the equatorial angular velocity at the bottom
of the convection zone and $\alpha_2$ is a parameter for the 
relative equator-to-pole difference of the angular velocity. 
Watson found instability for a differential rotation with 
$\alpha_2>0.286$. At that time, this was supposed to be in the vicinity
of the value of solar differential rotation, and the result was 
not definitely deciding between a turbulent and a stable tachocline. 
The investigation was two-dimensional arguing that the stable 
stratification will not allow significant radial flows, but 
thereby the radial shear is neglected, too.

Charbonneau et al. (1999a) extended the linear stability analysis
to various rotation profiles of the form
\begin{equation}
  \Omega = \Omega_{\rm eq}(1-\alpha_2 \cos^2\theta-\alpha_4 \cos^4\theta).
\end{equation}
Different layers of the tachocline were associated with different
results from helioseismology in terms of $\alpha_2$ and $\alpha_4$.
The upper layer which is thought to be penetrated by convective
overshooting was found to be unstable to the shear, whereas the
lower layer  with smaller latitudinal shear turned out to be 
stable. In a step towards the three-dimensional stability of the
tachocline, Dikpati \& Gilman (2001) studied the two-dimensional
system allowing for deformations in the third -- the radial --
dimension. As a result, the critical differential rotation for
instability was reduced to 11\% in the overshoot part of the
tachocline.


In order to address the entire tachocline and 
since the tachocline is a place where latitudinal and radial
shear meet, we investigate the stability of the three-dimensional
rotation profile. Although it is very reasonable to assume that radial
flows will be weak, the variation of the latitudinal differential
rotation with radius across weakly coupled spherical layers could
provide different results for the stability of the tachocline.
We will brief\/ly summarize the numerical background in 
Section~2, provide details of the computational results in Section~3
with different rotatation profiles,
and summarize our findings in Section~4. We find that the tachocline
is hydrodynamically stable in all the configurations studied. The
paper is thus intended to be a preparatory step towards fully 
three-dimensional studies of the MHD stability of the tachocline.

\section{Computational setup}
The rotational profile depends on both latitude and radius in 
this study. Between the inner and outer radius of the tachocline,
$r_{\rm i} = 0.65$ and $r_{\rm o}=0.7$ respectively, the angular velocity 
is defined by 
\begin{equation}
  \Omega(r,\theta) = \Omega_{\rm eq}\!\!\left[1-\alpha_2 \cos^2\theta-
  \alpha_2\left(\frac{1}{4}-\cos^2\theta\right)\!\frac{r_{\rm o}-r}{r_{\rm o}-r_{\rm i}}\right]\!\!,\,
  \label{omega}
\end{equation}
where $\theta$ denotes the colatitude in the spherical shell, $r$ is the 
radial coordinate, and $\Omega_{\rm eq}$ is the equatorial angular velocity.
The profile implies that the rotation velocity at the inner boundary
of the computational domain, $r=r_{\rm i}$, is the one at
$\theta=60^\circ$ (or $30^\circ$ heliographic latitude). This appears
to be a valid assumption in agreement with various helioseismological
inversions (recently e.g.\ Schou et al.\ 2002), whereas core rotation 
may be adopted at larger latitudes higher up in the convection zone.
The formation of the tachocline rotation profile is supposed to
be caused by rotating convection on top of it and reduction of differential
rotation by magnetic fields in the solar core at the bottom. Such a profile
causes a meridional circulation reaching a steady-state in competition with
the aforementioned (or other tachocline-forming) effects (see e.g.\ 
Sule et al.\ 2005). In our linear analysis, this flow has no effect on
the stability of the non-axisymmetric modes investigated in this Paper.

\begin{figure}
\epsfig{file=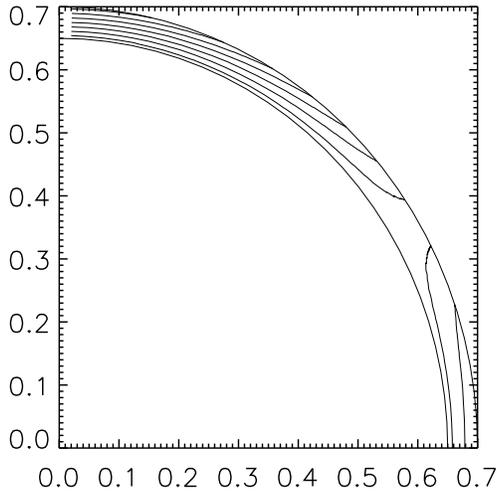,width=0.8\linewidth}
\caption{Vertical cross-section through the solar tachocline with
contours of the assumed angular velocity depending on radius and 
latitude as given by (2). Highest angular velocity is at the equator
(bottom right), and radially constant rotation occurs at a heliographic
latitude of $30^\circ$.}
\end{figure}

We employ the incompressible, viscous Navier-Stokes equation and linearize
the problem. We can then separate the axisymmetric background
rotation $\vec U$ from the nonaxisymmetric flow $\vec u$. The latter is evolved
by numerical computations. The normalised equation of motion reads
\begin{eqnarray}
\frac{\partial{\vec u}}{\partial t} &=&
  \vec u\times\nabla\times \vec U + \vec U\times\nabla\times \vec u\nonumber\\
& & -\nabla p-\nabla(\vec{u}\cdot \vec{U}) +\triangle\vec u,
\label{ns}
\end{eqnarray}
and the continuity equation $\nabla\cdot\vec u=0$ holds. The 
equation is evolved with the spectral spherical code by Hollerbach 
(2000). We look for exponentially decaying or growing solutions in
order to find the critical differential rotation $\alpha_2$ of the 
marginal case. The actual integration employs the radial components of
the curl and the curl-curl of the equation, thereby eliminating
the gradient terms. The linear (viscous) part of the full equation
of motion is evolved implicitly, while the nonlinear parts are integrated 
explicitly with the advection term and forces being computed in real
space. We kept this splitting even in our linearized problem,
since a fully implicit scheme would have required a substantial
modification of the code. The $\triangle\vec u$ is thus treated
implicitly, the other rhs terms (two remaining after curling)
are computed in real space and are used for a second-order Runge-Kutta 
integration.

The normalisation of the equation with the viscous time
$\tau_\nu = r_{\rm o}^2/\nu$ and the length scale $r_{\rm o}$
leads to the Reynolds number
\begin{equation}
{\rm Re} = \frac{r_{\rm o}^2\Omega_{\rm eq}}{\nu}
\label{Re}
\end{equation}
as a free parameter which is essentially a variation of the
viscosity $\nu$ since radius and $\Omega_{\rm eq}$ are sufficiently
well known. The solar Reynolds number in the tachocline is
-- in terms of the definition of (\ref{Re}) -- about $10^{14}$. 
We try to achieve time series for numerically
demanding ${\rm Re}>10^4$. By comparison with known results from
inviscid two-dimensional analyses, we find that the critical viscous
differential rotation at ${\rm Re}>10^3$ or $10^4$ is already 
sufficiently close to the inviscid value.

The velocity is decomposed into toroidal and poloidal potentials, 
$\vec u = \nabla\times(e\hat{\vec r}) + \nabla\times\nabla\times(f\hat{\vec r})$,
where $\hat{\vec r}$ is the unit vector in radial direction.
The potentials are again decomposed into radial Chebyshev polynomials 
and spherical harmonics. Since it is the potentials being evolved, the
continuity equation is fulfilled automatically. The density is
constant throughout the computational domain. In a thin shell 
of 5\% of the solar radius, this is a reasonably good approximation
of the true situation. We also do not take any deformation of the
tachocline into account. Top and bottom radius of the tachocline
are constant over latitude. Charbonneau et al. (1999b) found a
prolate tachocline whose equatorial part is located at slightly 
smaller radius than the polar end. We assume that the difference
of 3.5\% in location of the tachocline has negligible effect
on the results, but may be an issue of future investigations.

The radial boundary conditions for the velocity perturbations
are stress-free at both $r_{\rm i}$ and $r_{\rm o}$. At Reynolds
numbers of ${\rm Re}>10^4$, high spectral resolution was
necessary to obtain reliable results. Up to 80 Chebyshev
and 80 Legendre polynomials were used to resolve the flow
properly.

The azimuthal modes of the problem described by 
(\ref{ns}) are decoupled, and we can study the
stability of individual $m$-modes separately. Moreover,
even and odd latitudinal modes (symmetric and antisymmetric
modes with respect to the equator) decouple, and we will
have a look into the critical differential rotation for
the excitation of instability of the two kinds separately.
The radial modes all couple and do not provide results on
the stability of individual radial wavelengths.

\begin{figure}
\centerline{
\epsfig{file=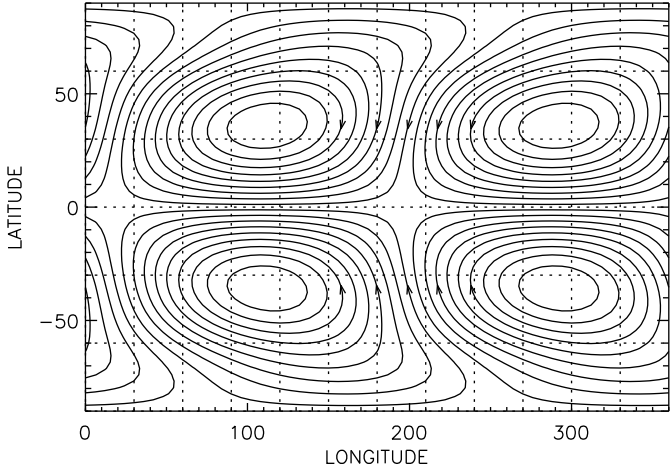,width=\linewidth}}
\caption{Streamlines of the symmetric eigen function of the computation
with a three-dimensional profile $\Omega(r,\theta)$ on the 
surface of the tachocline at $r=0.7$. Two circulation cells 
are found on each hemisphere.\label{potsurf}}
\end{figure}

\section{Results}
\subsection{Stability of various solutions}
In a set of fiducial computations, we applied a purely latitudinal
profile of the angular velocity. An $m=1$ mode is evolved with 
the profile of (1) where $\alpha_4=0$ and $\alpha_2$ is
varied. Since we solve a viscous problem,
the critical differential rotation, $\alpha_2^{\rm crit}$,
depends on the Reynolds number. The result is already very
close to Watson's inviscid result for ${\rm Re}\geq 1000$.
This is good reason to assume that the numerical solutions
reaching ${\rm Re}\sim 30\,000$ are suitable approximations
for the solar plasma. 

Despite allowing for radial motions, the evolution provides
solutions which are nearly toroidal and do not show significant 
radial flows. They are surface flows forming two cells on 
each hemisphere with stream lines through the poles. 
Figure~\ref{potsurf} shows a representation  of the
flow in the spherical surface. The graph has to assume that
the poloidal component of the velocity is zero though,
which is not entirely true. 

\begin{figure}
\epsfig{file=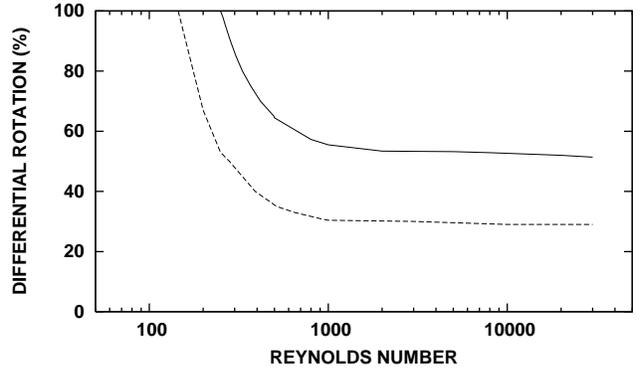,width=\linewidth}
\caption{Lines of marginal stability for the combined latitudinal
and radial shear (solid line) and the purely latitudinal
shear (dashed line). Differential rotation denotes the
percentage by which the pole's angular velocity is slower 
compared with the equatorial one. It is expressed by $\alpha_2$
from (1) and (2) here.\label{stability}}
\end{figure}

The second step involved the rotation profile of (2)
for which the critical steepness of the differential rotation, 
$\alpha_2^{\rm crit}$, is again sought for various Reynolds
numbers. Figure~\ref{stability} shows the lines of marginal stability,
i.e. the critical differential rotation, versus Reynolds number for the
symmetric $m=1$ mode. The solid line refers to profile~(2), 
the dashed line is the latitudinal 
profile and converges to the result by Watson (1981) for
${\rm Re}\rightarrow\infty$. 

\begin{figure}
\epsfig{file=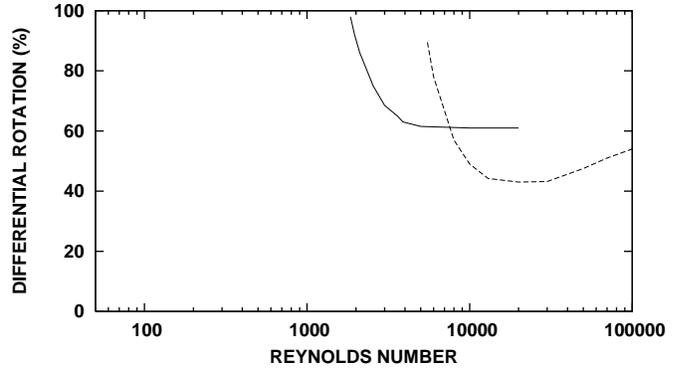,width=\linewidth}
\caption{Lines of marginal stability for the antisymmetric $m=1$
solutions caused by the combined latitudinal
and radial shear (solid line) and the purely latitudinal
shear (dashed line).\label{antisym}}
\end{figure}

The most easily excited patterns of $m=1$ are always 
symmetric with respect to the equator.
We can also look for the stability of antisymmetric patterns
and find the results shown in Fig.~\ref{antisym}. They are more
stable than the symmetric configurations with an 
$\Omega(r,\theta)$ profile. The antisymmetric solutions
from the $\Omega(\theta)$ profile have also higher 
critical differential rotation values than their
symmetric counterparts.

\begin{figure}
\epsfig{file=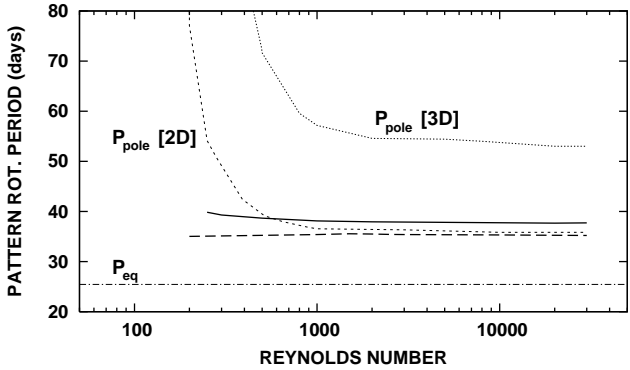,width=\linewidth}
\caption{Rotation period of the flow pattern for the two-dimensional
(long-dashed) and the three-dimensional (solid) rotation profile. The
numbers are computed assuming an equatorial rotation period of 
25.44~d ($\Omega_{\rm eq}=455$~nHz) which is plotted with a dash-dot
line. Periods are computed at marginal stability; the polar rotation
period for this critical differential rotation is plotted as
short-dashed and dotted lines for the 2D and 3D cases, respectively.\label{period}}
\end{figure}

The patterns drift with a certain velocity in azimuthal direction 
around the solar axis. Since the equations hold for the nonrotating
system, we can directly convert the pattern rotation
into physical times. The pattern rotation period for
the two-dimensional and the three-dimensional profile
of $\Omega$ is shown in Fig.~\ref{period} by a dashed and a solid
line, respectively. The actual solar rotation periods are also 
plotted. The reference equatorial period of 25.44~d
($\Omega_{\rm eq}=455$~nHz) is given as dash-dot line.
The pattern rotation periods are determined at marginal
stability. Since the marginal case gives us a value for
the differential rotation, we also plot the polar
rotation period $P_{\rm pole}$ versus Re. The short-dash
line is for the two-dimensional $\Omega(\theta)$ profile, the dotted line
is for the three-dimensional case described by (2).

The pattern rotation periods are always between the equatorial
and polar rotation periods, in agreement with the 2D results
by Charbonneau et al. (1999a). While the patterns from the 
2D-$\Omega$ profile are close to the polar rotation period,
the patterns for the 3D profile rotate with nearly the average rotation 
period between the polar and equatorial ones. 

We can compute the time
after which the pattern is passed by a given point on the equator.
This time is often called lap time. Assuming an equatorial 
rotation period of 25.44~d ($\Omega_{\rm eq}=455$~nHz),
we find a lap time of 91~d for the 2D case, and a lap
time of 78~d for the 3D case. 

\begin{figure}
\epsfig{file=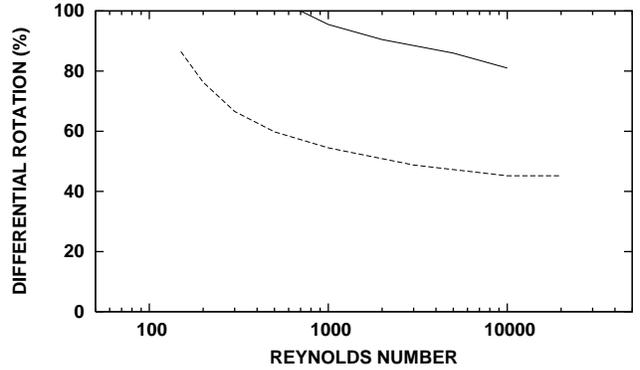,width=\linewidth}
\caption{Lines of marginal stability for the antisymmetric $m=2$ mode resulting
from the combined latitudinal and radial shear (solid line) 
and the purely latitudinal shear (dashed line).\label{stability_m2}}
\end{figure}

Modes with higher azimuthal mode numbers require significantly
higher differential rotation for instability. The antisymmetric $m=2$ 
mode which is symmetric with respect to the equator was found to be
stable even in the entire parameter range covered by Fig.~\ref{stability}. 
This is in agreement with the inviscid, two-dimensional stability
analysis by Charbonneau et al. (1999a). The stability lines for the 
antisymmetric $m=2$ mode are shown in Fig.~\ref{stability_m2}. We could not find
instability for any $m=3$ mode in the range covered by Fig.~\ref{stability}.

\subsection{Effects of buoyancy}
Little influence is expected from the stable temperature gradient
in the tachocline. Since we do have the chance to prove this in
our three-dimensional simulations, we demonstrate the effect of
a negative buoyancy force on the stability of the differential
rotation. The Navier-Stokes equation is extended by the buoyancy
force and reads in the non-dimensional form:
\begin{eqnarray}
\frac{\partial{\vec u}}{\partial t} &=&
  \vec u\times\nabla\times \vec U + \vec U\times\nabla\times \vec u\nonumber\\
& & +{\rm Ra} \Theta \vec r -\nabla p-\nabla(\vec{u}\cdot \vec{U}) +\triangle\vec u,\\
\frac{\partial\Theta}{\partial t} &=&
  -\vec U\cdot \nabla\Theta - \vec u\cdot\nabla T + \frac{1}{{\rm Pr}}\triangle\Theta,
\label{nstemp}
\end{eqnarray}
with a background temperature profile of
\begin{equation}
  T=\frac{r_{\rm i}}{r_{\rm o}-r_{\rm i}}\left(\frac{r_{\rm o}}{r}-1\right)
\end{equation}
and the Prandtl number being the ratio between viscosity and
thermal diffusivity, ${\rm Pr}=\nu/\chi$. The Rayleigh number
in (\ref{nstemp}) is
\begin{equation}
 {\rm Ra} = \frac{g\alpha(T_{\rm i}-T_{\rm o})r_{\rm o}^3}{\nu^2},
\end{equation}
where $g$ is the gravitational acceleration, $\alpha$ is the 
coefficient of volume expansion, and $T_{\rm i}$ and $T_{\rm o}$ 
are the temperatures at the two boundaries. In the Boussinesq
formulation used here, the presence of a sub-adiabatic temperature
gradient actually translates into a negative value of Ra. We 
set our version of the Rayleigh number to a value as small 
(``as negative'') as ${\rm Ra}=-10^8$ in order to see any notable 
effect on the flow. The Prandtl number is set to unity.

The critical differential rotation for a growing symmetric
$m=1$ mode at ${\rm Re}=10^4$ increases slightly to 53.3\%
as compared with the non-buoyant value of 52\%. This is 
in line with the fact that the solutions contain nearly
horizontal motions.

\begin{figure}
\epsfig{file=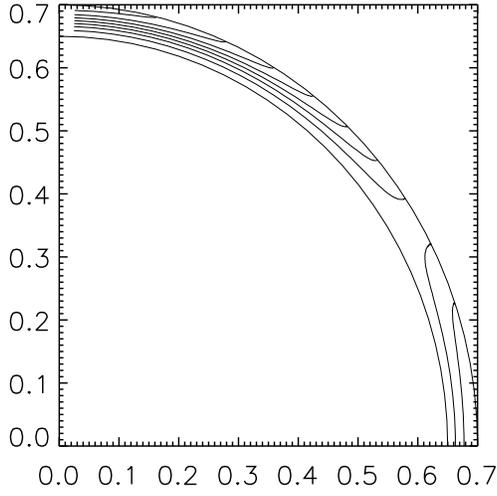,width=0.8\linewidth}
\caption{Contours of the assumed angular velocity with a radial
dependence as in (\ref{omega2}). Highest angular velocity is again
at the equator (bottom right), and radially constant rotation 
occurs at a heliographic latitude of $30^\circ$.\label{sps_omega2}}
\end{figure}

\subsection{Effects of higher-degree terms}
The differential rotation has been expressed by a $\cos^2\theta$ dependence
in latitude and a linear $r$ dependence over radius. We are studying the
stability of the three-dimensional tachocline setup with higher-degree
dependences in this Section, such as the $\cos^4\theta$ term and
a nonlinear radial dependence of $\Omega$.

Including the $\cos^4\theta$ term in (2) yields an angular velocity
profile of the form
%
%
\begin{eqnarray}
  \Omega(r,\theta) &=& \Omega_{\rm eq}\biggl\{ 1-\alpha_2 \cos^2\theta-\alpha_4 \cos^4\theta-
  \frac{r_{\rm o}-r}{r_{\rm o}-r_{\rm i}}\cdot\nonumber\\
  &&\left.\cdot\left[\alpha_2\left(\frac{1}{4}-\cos^2\theta\right)
  +\alpha_4\left(\frac{1}{16}-\cos^4\theta\right)\right]\right\}
\end{eqnarray}
The computations for the full profile could not easily be extended 
beyond ${\rm Re}=5000$. But the results for the possible Re and for
the easier two-dimensional $\Omega(\theta)$ show no decreased
critical differential rotation when the $\cos^4\theta$ term is
included. This also holds for the extreme case of $\alpha_4$
carrying the shear alone ($\alpha_2=0$).

%
%
Starting from (\ref{omega}), a modified $\Omega$-profile was constructed in order to find
any influence of the particular radial dependence of the latitudinal
shear on the results. We used
\begin{eqnarray}
  \Omega(r,\theta) &=& \phantom{-}\Omega_{\rm eq}\biggl[1-\alpha_2 \cos^2\theta-\nonumber\\
  &&\left.-\alpha_2\left(\frac{1}{4}-\cos^2\theta\right)
  \!\frac{1}{2}\!\left(1-\cos\Bigl(\frac{r_{\rm o}-r}{r_{\rm o}-r_{\rm i}}\pi\Bigr)\!\!\right)\right]
  \label{omega2}
\end{eqnarray}
The radial profile introduces two inflection points. It does not
apply the usual error function for the radial $\Omega$-step in
order to obtain an exact $\partial\Omega/\partial r = 0$ at 
both inner and outer boundary. A graphic representation is shown 
in Fig.~\ref{sps_omega2}.

\begin{figure}
\epsfig{file=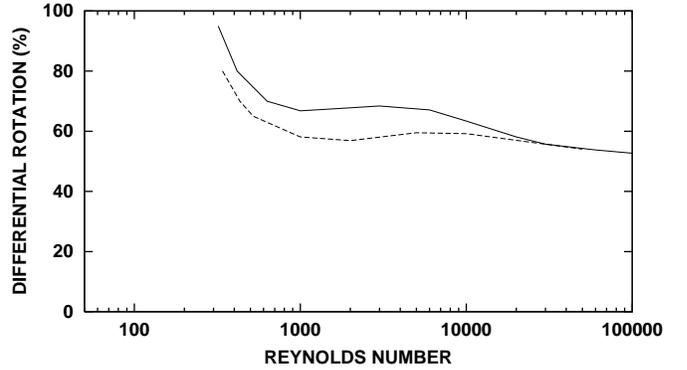,width=\linewidth}
\caption{Lines of marginal stability for an angular
velocity profile given by (\ref{omega2}) having two
inflection points over radius. The solid line refers to
the non-buoyant case, ${\rm Ra = 0}$, the dashed line
has ${\rm Ra} = -10^8$. \label{stability_inflection}}
\end{figure}

The critical differential rotation for instability is shown
versus Reynolds number in Fig.~\ref{stability_inflection}.
We were able to reach extraordinary Reynolds numbers of 
$10^5$, probably because of the vanishing $\partial\Omega/\partial r$
at the boundaries. Instability does not occur at reduced differential 
rotation, and the line of marginal stability actually 
converges to the simpler profile (2) for high Re.
The additional effect of buoyancy on the stability of the
profile (\ref{omega2}) with inflection points is also shown as 
a dashed line.

\section{Summary}
A fully three-dimensional, linear analysis of the stability
of the solar tachocline was carried out. If radial variation
of the angular velocity is included in the model, the maximum
pole-equator difference of the angular velocity can be as 
large as 52\% for a symmetric $m=1$ mode, before instability
sets in. Two-dimensional analyses have delivered much 
lower critical values. The difference of 3D versus 2D
is not radial flows emerging from the extension
in the third dimension, but it is changed stability conditions
emerging from the radial shear and radial dependence of the
differential rotation.

Other modes such as higher $m$ or different flow symmetries
do not get unstable at lower critical differential rotation
values under the influence of a three-dimensional rotation
profile.

The stabilizing effect of the temperature gradient has been
added, but since all the unstable modes are very nearly
horizontal, the influence is small. The assumption that 
horizontal motions dominate is valid even without a stabilizing 
temperature gradient. However the assumption that spherical
shells of infinitesimal thickness do not interact with each 
other is not applicable, according to our results. One may
argue that the viscosity in the computer simulations is
much too high, but the variation of the results is small
at ${\rm Re}>1000$. This is an indication that computations
with ${\rm Re}=10^4$ or higher are a good approximation of
the near-inviscid solar case. We conclude that all parts
of the tachocline not being affected by convective overshooting
are stable. We believe that the answer on the question of the 
dynamical state of the tachocline will be given by MHD computations
of a three-dimensional domain.

\begin{acknowledgements}
We are grateful to Peter Gilman for his valuable comments on
the topic. AS thanks the Deutsche Forschungsgemeinschaft
for the support by grant No.~Ru\,488/15-1.
\end{acknowledgements}

\end{document}